\documentclass[aps,prl,showpacs,twocolumn,superscriptaddress,longbibliography,nofootinbib]{revtex4-1}

\usepackage[english]{babel}

\usepackage{relsize}
\usepackage{longtable}
\usepackage{booktabs}

\usepackage{graphicx}
\usepackage{amsmath}
\usepackage{multirow}
\usepackage{xcolor}

\def\aap{Astron. Astrophys.} 
\def\mnras{Mon. Not. R. Astron. Soc.}           
\def\pasp{Publ. Astron. Soc. Pac.}               
\def\aj{A.J.}

\begin{document}


\title{Rotational and rotational-vibrational Raman spectroscopy of air to characterize astronomical spectrographs}


\author{Fr\'ed\'eric P.~A. Vogt}
\email[]{frederic.vogt@alumni.anu.edu.au}
\thanks{ESO Fellow}
\affiliation{European Southern Observatory (ESO), Av. Alonso de C\'ordova 3107, 763 0355 Vitacura, Santiago, Chile.}

\author{Florian Kerber} 
\affiliation{European Southern Observatory (ESO), Karl-Schwarzschild-Str. 2, 85748 Garching, Germany.}
\author{Andrea Mehner}  
\affiliation{European Southern Observatory (ESO), Av. Alonso de C\'ordova 3107, 763 0355 Vitacura, Santiago, Chile.}

\author{Shanshan Yu} 
\affiliation{Jet Propulsion Laboratory, California Institute of Technology, Pasadena, CA, 91109, USA.}

\author{Thomas Pfrommer} 
\affiliation{European Southern Observatory (ESO), Karl-Schwarzschild-Str. 2, 85748 Garching, Germany.}

\author{Gaspare Lo Curto} 
\affiliation{European Southern Observatory (ESO), Av. Alonso de C\'ordova 3107, 763 0355 Vitacura, Santiago, Chile.}

\author{Pedro Figueira}  
\affiliation{European Southern Observatory (ESO), Av. Alonso de C\'ordova 3107, 763 0355 Vitacura, Santiago, Chile.} 
\affiliation{Instituto de Astrof\'isica e Ci\^encias do Espa\c co, Universidade do Porto, CAUP, Rua das Estrelas, 4150-762 Porto, Portugal.}

\author{Diego Parraguez} 
\affiliation{European Southern Observatory (ESO), Av. Alonso de C\'ordova 3107, 763 0355 Vitacura, Santiago, Chile.}

\author{Francesco A. Pepe} %
\author{Denis M\'egevand}   
\affiliation{Observatoire Astronomique de l'Universit\'e de Gen\`eve, 51 chemin des Maillettes, 1290 Versoix, Switzerland.}

\author{Marco Riva} %
\affiliation{Osservatorio Astronomico di Brera, INAF, Via Bianchi 46, 23807 Merate, Italy.}

\author{Paolo Di Marcantonio} 
\affiliation{Osservatorio Astronomico di Trieste, INAF, Via Tiepolo 11, 34143 Trieste, Italy.}

\author{Christophe Lovis} 
\affiliation{Observatoire Astronomique de l'Universit\'e de Gen\`eve, 51 chemin des Maillettes, 1290 Versoix, Switzerland.}

\author{Manuel Amate} %
\affiliation{Instituto de Astrof\'isica de Canarias, C/. V\'ia L\'actea, 38200 La Laguna, Tenerife, Spain.}

\author{Paolo Molaro} 
\affiliation{Osservatorio Astronomico di Trieste, INAF, Via Tiepolo 11, 34143 Trieste, Italy.}

\author{Alexandre Cabral} 
\affiliation{Faculdade de Ci\^{e}ncias, Instituto de Astrof\'isica e Ci\^encias do Espa\c co, Universidade de Lisboa, Campo Grande, Edif\'icio C8, 1749-016 Lisboa, Portugal.}

\author{Maria Rosa Zapatero Osorio} 
\affiliation{Centro de Astrobiolog\'ia (CSIC-INTA), Carretera de Ajalvir km 4, Torrej\'on de Ardoz, 28850 Madrid, Spain.}


\date{\today}

\begin{abstract}
Raman scattering enables unforeseen uses for the laser guide-star system of the Very Large Telescope. Here, we present the observation of one up-link sodium laser beam acquired with the ESPRESSO spectrograph at a resolution $\lambda/\Delta\lambda \sim 140'000$. In 900\,s on-source, we detect the pure rotational Raman lines of $^{16}$O$_2$, $^{14}$N$_2$, and $^{14}$N$^{15}$N (tentatively) up to rotational quantum numbers $J$ of 27, 24, and 9, respectively. We detect the $^{16}$O$_2$ fine-structure lines induced by the interaction of the electronic spin \textbf{S} and end-over-end rotational angular momentum \textbf{N} in the electronic ground state of this molecule up to $N=9$. The same spectrum also reveals the $\nu_{1\leftarrow0}$ rotational-vibrational Q-branch for $^{16}$O$_2$ and $^{14}$N$_2$. These observations demonstrate the potential of using laser guide-star systems as accurate calibration sources for characterizing new astronomical spectrographs. 
\end{abstract}


\maketitle


\section{Introduction}
The 4 Laser Guide Star Facility \citep[4LGSF;][]{BonacciniCalia2014} has been in operation on Unit Telescope 4 (UT4) of the Very Large Telescope (VLT) in Chile's Atacama desert since mid-2016. It is comprised of four 22\,W continuous wave lasers, and forms an integral part of UT4's Adaptive Optics Facility \citep[AOF;][]{Arsenault2013}. Its lasers are used to excite sodium atoms in the mesosphere, primarily located in a layer several kilometers thick at an altitude of $\sim$90\,km \citep{Moussaoui2010,Neichel2013,Pfrommer2014}. The four lasers each emit 18\,W at 5891.59120\,\AA\ to excite the D$_2$a sodium transition, 2\,W at 5891.57137\,\AA\ to re-pump (via the D$_2$b line) sodium atoms ``lost'' to the 3$^2S_{1/2}$ F=1 state \citep[with F the total atomic angular momentum quantum number; see][]{Holzlohner2010}, and 2\,W at 5891.61103\,\AA\ that are not usable for adaptive optics (AO) purposes. The spectral stability of the lasers, of the order of $\pm$3\,MHz$\equiv$3.5\,fm over hours, is achieved by using a solid-state high-resolution wavelength meter. The absolute accuracy of the lasers ($<$10\,MHz$\equiv$11.6\,fm at the 3$\sigma$ level) is achieved via periodic calibrations of the wavelength meter against a stabilized Helium-Neon reference laser \citep{Friedenauer2012,Enderlein2014, Lewis2014}. The resulting four wavefront reference sources created in the mesosphere are exploited by the AO modules GALACSI \citep{Stuik2006,LaPenna2016} and GRAAL \citep{Paufique2010} that are coupled to the astronomical instruments MUSE \citep{Bacon2010} and HAWK-I \citep{Kissler-Patig2008,Siebenmorgen2011}, respectively.

The 4LGSF was designed to work in concert with the other components of the AOF to enhance the image quality achieved by instruments on UT4. The MUSE observations of the inelastic Raman scattering of laser photons by air molecules above UT4 \citep{Vogt2017b} have, however, opened unforeseen avenues for experimentation with the 4LGSF system in a \textit{stand-alone} mode\footnote{By \textit{stand-alone}, we mean that the 4LGSF is operated on its own, as opposed to the \textit{nominal} mode in which the 4LGSF is slaved to the the other components of the AOF to support AO observations with MUSE and HAWK-I.}. For example, Telescope and Instrument Operators can easily vary the size of the square asterism of laser guide-stars by using the Laser Pointing Camera \citep[][]{BonacciniCalia2014a}, without the need to interact with any of the other AOF components or UT4 instruments. This capability was used to monitor the flux of laser lines present in MUSE AO observations over a 27-months period, which revealed dust particles on the primary and tertiary telescope mirrors to be an important secondary scattering source for the laser photons \citep{Vogt2018c}. Here, we discuss another stand-alone use-case for the 4LGSF, and laser guide-star systems in general: that of an accurate wavelength calibration source for astronomical spectrographs.

The Echelle SPectrograph for Rocky Exoplanets and Stable Spectroscopic Observations \citep[ESPRESSO;][]{Pepe2010,Pepe2013} is an ultra-stable fibre-fed high-resolution spectrograph installed in the Coud\'e facility of the VLT. The instrument delivers a spectral resolution $\lambda/\Delta\lambda$ of $\sim$140'000 (single UT, ``HR'' mode), $\sim$190'000 (single UT, ``UHR'' mode), or $\sim$70'000 (four UTs, ``MR'' mode). The spectrograph itself is temperature-stabilized at the mK level in a 10$^{-5}$ hPa vacuum. Astronomical observations can be wavelength-calibrated with a Th-Ar hollow-cathode lamp (that can be aided by a white-light Fabry-P\'erot), or a dedicated Laser Frequency Comb \citep{LoCurto2012}. In particular, the existence of two parallel fibers allows to acquire scientific observations (in fibre ``A'') with a simultaneous wavelength calibration exposure (in fibre ``B'') to track internal radial velocity drifts over time. The overall instrument design is driven by the goal \textit{precision} of 10\,cm\,s$^{-1}$ over 10 years for radial-velocity measurements in a single UT mode. The goal wavelength \textit{accuracy} of ESPRESSO, on the other hand, is 10\,m\,s$^{-1}$.

As with all new systems at the VLT, ESPRESSO underwent a series of commissioning observations to characterize its performances as-built, following its installation at the Coud\'e focus, and prior to being offered to the general community. Characterizing the wavelength calibration accuracy of the spectrograph is evidently one of the high-priority goals for this activity. Ideally, this should be achieved by means of external reference sources that fully mimic real observations, and allow to characterize the entire \textit{telescope(s), Coud\'e train, and instrument} assembly. The 4LGSF, and the associated Raman scattering of its laser photons by air molecules, provides the ideal means to do so. In this Letter, we present the first ESPRESSO observation of one 4LGSF up-link laser beam acquired during the commissioning phase of the instrument. Throughout the text, all quoted wavelengths are in vacuum.

\section{Observations and data reduction} \label{sec:obs}
\begin{figure}[htb!]
\vspace{25pt}
\centerline{\includegraphics[scale=0.5]{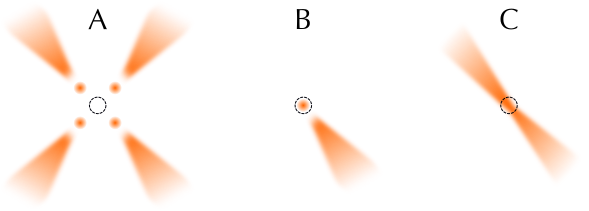}}
\caption{Diagram (not to scale) of the manual acquisition steps for the ESPRESSO observations described in this article. A: all 4LGSF lasers are propagating. The 4 laser guide-stars (orange dots) are positioned to a nominal square asterism using the LPC. From the instrument perspective, the up-link laser beams (visible up to $\sim$35\,km of altitude) are located on the outside of the laser guide-stars (located at $\sim$90\,km of altitude). B: using the LPC, a single laser guide-star is placed in centerfield. The propagation is stopped for all other laser guide-stars. C: the up-link beam is brought over the ESPRESSO fiber (black circle) using manual offsets applied to the jitter-loop mirror of the laser launch telescope. The focus of the telescope is adjusted to bring the beam better into focus at the location of the fiber.}\label{fig:acq}
\end{figure}

The ESPRESSO spectrum presented in this article was acquired during commissioning activities on the night of February 2, 2018. The observations were performed from UT4, with the 4LGSF system operated in stand-alone mode, in parallel to ESPRESSO which was entirely oblivious to it. The telescope was preset to place the ESPRESSO fibre A on an empty sky field, with no entries in the USNO-B1 catalogue \citep[which is complete down to V=21 mag;][]{Monet2003}. Upon the end of the acquisition sequence, one 22\,W laser guide-star was placed manually in the center of the VLT field-of-view using the Laser Pointing Camera (LPC; see Fig.~\ref{fig:acq}). Manual offsets were then applied to the jitter-loop mirror of the laser launch telescope to bring the up-link beam -- clearly visible in the ESPRESSO Technical CCD of Front-End 4 for Field Acquisition \citep{Duhoux2014} -- into the field-of-view of the ESPRESSO fiber. The propagation of the other three laser guide-stars was stopped. The spectrum presented here corresponds to a single 900\,s exposure, acquired in the HR21 mode ($\lambda/\Delta\lambda\sim140'000$, 1$^{\prime\prime}$ fibre diameter on-sky, $2\times1$ pixel binning along the ``spatial'' direction). The observation was performed at an airmass of 1.01 (equivalent to a telescope altitude of 81.7\,deg at the start), with 16\,mm of precipitable water vapor \citep[which is a very high value for Cerro Paranal which has a median of $\sim$2.4\,mm;][]{Kerber2012,Kerber2014}, a relative ambient humidity of 38\%, an ambient pressure of 742.4\,hPa, and a ground temperature of 13.5\,$^{\circ}$C. The exact beam altitude probed is formally unknown: from our experience with MUSE \citep[see the Appendix in][]{Vogt2017b}, we estimate it to be $15\pm8$\,km above ground. Given the perspective and the continuous-wave nature of the 4LGSF lasers, one has to note that a range of several km is being sampled by the ESPRESSO fiber. Since the laser beam was not located at infinity, the secondary mirror of UT4 was offset by +4\,mm to bring the beam into better focus, and thus increase the observed beam surface brightness.

The data were reduced using the ESPRESSO pipeline v1.1.4. The wavelength calibration was derived from reference exposures (in fiber A) obtained with the Laser Frequency Comb during the daily instrument calibrations.

\newpage

\section{Results}\label{sec:results}

\begin{figure*}[htb!]
\centerline{\includegraphics[scale=0.5]{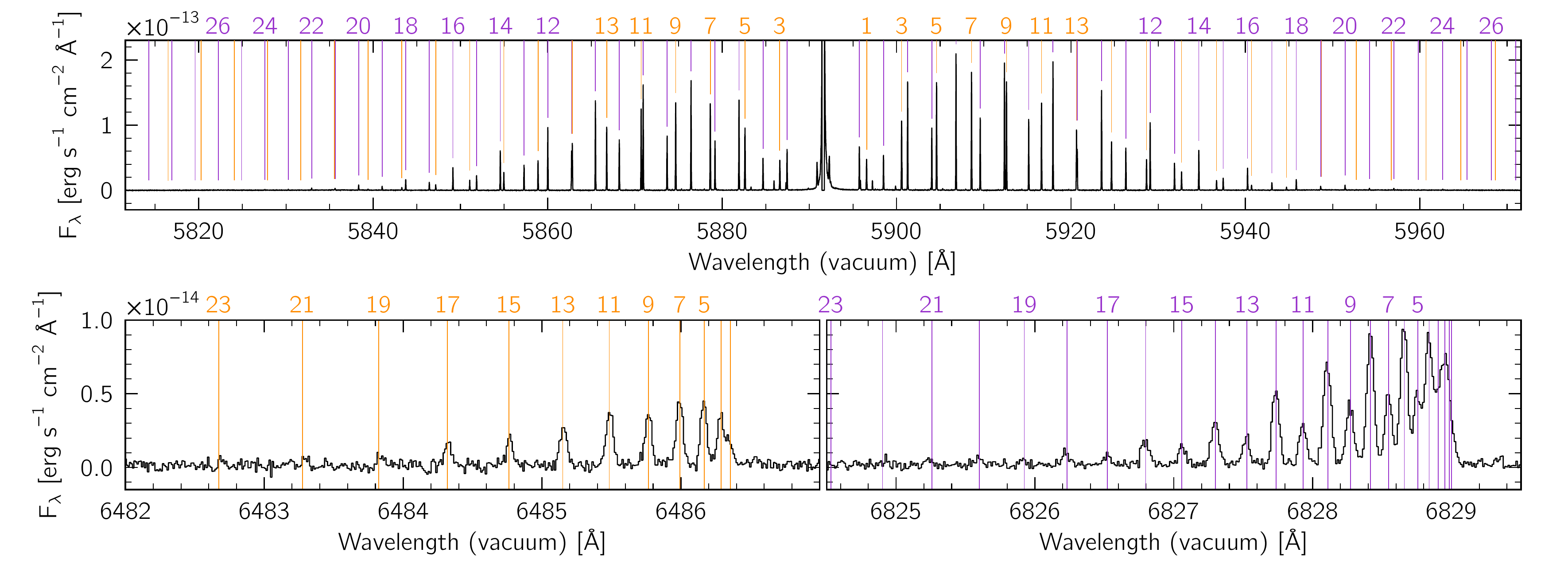}}
\caption{ESPRESSO view of the pure rotational Raman lines (top), and $Q$ branch of the ro-vibrational Raman lines of $^{16}$O$_2$ (bottom left) and $^{14}$N$_2$ (bottom right), associated with one 4LGSF up-link laser beam. Orange (for $^{16}$O$_2$) and purple (for $^{14}$N$_2$) lines mark the theoretical position of each Raman line derived with the model of non-rigid singlet diatomic rotators (see the Supplemental Material for details). Colored integers denote the rotational quantum number  $J$ associated with the different analytical wavelength predictions. The CCD did saturate within $\pm$0.2\,\AA\ of the main laser line.}\label{fig:rovib}
\end{figure*}

\begin{figure*}[htb!]
\centerline{\includegraphics[scale=0.5]{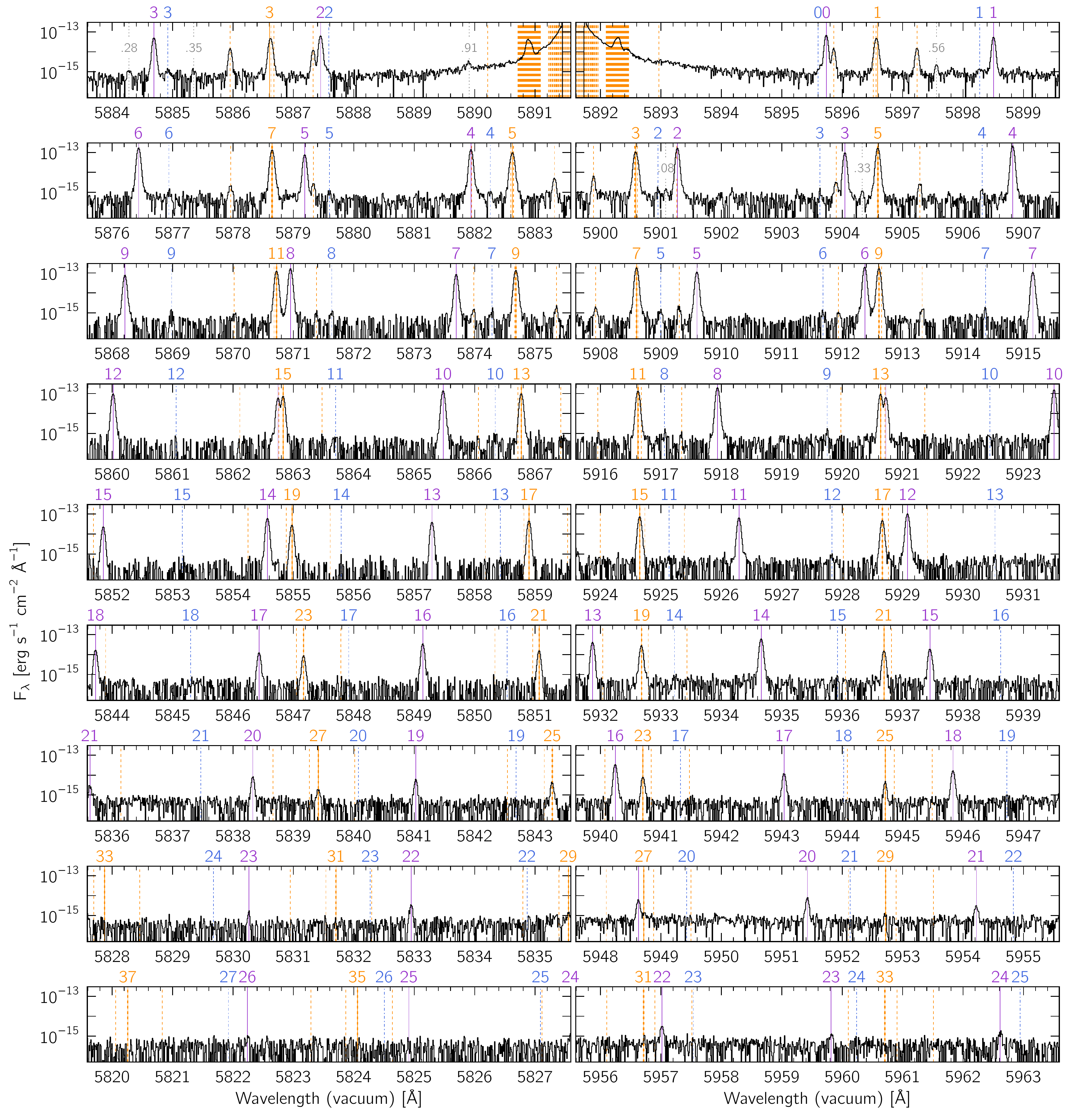}}
\vspace{-10pt}
\caption{ESPRESSO spectrum of one 4LGSF up-link laser beam, focusing on the spectral regions containing pure rotational Raman lines. Each panel is 8\,\AA\ wide. For a given row, the left and right panels are equidistant from the main laser line at 5891.5912\,\AA\ (which is saturated). Pure rotational Raman lines, associated with the Stokes (right column) and anti-Stokes (left column) branches, are detected up to $J=27$ for $^{16}$O$_2$, $J=24$ for $^{14}$N$_2$, and $J=9$ for $^{14}$N$^{15}$N. Solid orange, solid purple, and dash-dotted blue vertical lines mark the theoretical position of the pure rotational Raman lines of $^{16}$O$_2$, $^{14}$N$_2$, and $^{14}$N$^{15}$N (respectively), when treated as non-rigid singlet diatomic rotators. The molecular rotational quantum number $J$ associated with each of these lines is indicated above them (only the $^{14}$N$_2$ $J=11$ anti-Stokes and $J=9,19$ Stokes lines are not tagged because of crowding). Dashed orange vertical lines, derived using a numerical model of $^{16}$O$_2$ accounting for the non-zero electronic spin of its ground state, highlight the fine-splitting of the pure rotational lines for this molecule. The nature of 6 lines (tagged with dotted grey vertical lines and the first two decimals of their wavelength in the top panels) remains unknown at this time.}\label{fig:rot}
\end{figure*}

We present in Fig.~\ref{fig:rovib} and \ref{fig:rot} a subset of the ESPRESSO spectrum of one up-link laser beam from the 4LGSF. Fig.~\ref{fig:rovib} contains an overall view of the pure rotational Raman lines and the $\nu_{1\leftarrow0}$ rotational-vibrational (ro-vibrational) Raman lines for $^{16}$O$_2$ and $^{14}$N$_2$, readily visible as a resolved forest of lines. In Fig.~\ref{fig:rot}, we present a detailed view of the spectral regions located within $\pm$70\,\AA\ from the main laser line. To identify the specific origin of each line in the spectrum, we first compare it with the analytical predictions of the rotational and ro-vibrational Raman lines for homonuclear diatomic molecules treated as non-rigid singlet rotators \citep[see the Supplemental Material for the full analytical derivation, which includes Refs.][]{Herzberg1950,Shimauchi1995,Orlov1997,Bendtsen2001,Huber1979}. We unambiguously detect pure rotational lines up to a molecular rotational quantum number $J=27$ for $^{16}$O$_2$ and $J=24$ for $^{14}$N$_2$. The lines in the $\nu_{1\leftarrow0}$ ro-vibrational Q-branch of $^{14}$N$_2$ and $^{16}$O$_2$ are resolved from $J\gtrsim4$, and unambiguously detected up to $J\sim18$. For $^{16}$O$_2$, only lines associated with odd values of $J$ are detected, as expected from the selection rules imposed by quantum mechanics for homonuclear diatomic molecules with zero nuclear spin \citep{Herzberg1950}. 

We also detect the pure rotational Raman lines from $^{14}$N$^{15}$N (tentatively) up to $J=9$. These lines are detected with a signal-to-noise ratio $\cong2\pm1$: insufficient to formally rule out $^{14}$N$_2$$^+$ as the molecule responsible, from the theoretical line wavelengths alone (see the Supplemental Material for details). To try to discriminate between $^{14}$N$^{15}$N and $^{14}$N$_2^{+}$, we assembled a simple model of two Gaussian lines with their dispersion tied, to individually fit each of the tentative $^{14}$N$^{15}$N lines alongside the nearest $^{14}$N$_2$ line (i.e. associated to the same $J$-value). From a Markov-Chain Monte-Carlo sampling of the individual posterior distribution of each of the clearest 14 line pairs, assuming a least-square likelihood and flat priors with a lower bound of 0 for the line intensities and dispersion, we find no evidence for an alternating intensity pattern between lines associated with even and odd $J$ values, as one would expect from $^{14}$N$^{+}$. On the other hand, we derive an overall line intensity ratio of 0.5$^{+0.6}_{-0.3}$\% (68\% confidence level) with respect to the nearest $^{14}$N$_2$ lines, consistent with the atmospheric abundance ratio of $^{14}$N$^{15}$N to $^{14}$N$_2$ \citep{Junk1958}. Clearly, deeper observations are necessary to confirm this identification.

Altogether, the analytical predictions account for most of the rotational and all of the ro-vibrational lines visible in the ESPRESSO spectrum, but fail to explain a series of fainter lines in the vicinity of the main laser line. We link the majority of these fainter lines to the existence of a non-zero electronic spin in the ground state $^3\Sigma_g^-$ of $^{16}$O$_2$. It is the interaction between this electronic spin and the molecular rotation that leads to the fine-splitting of the pure rotational Raman lines of $^{16}$O$_2$. From an exhaustive numerical modeling of this effect \citep[see the Supplemental Material for details, which includes Refs.][]{Yu2014,Yu2012,Drouin2012,Drouin2013}, we can successfully identify side-lines up to $N=9$ (where $N$ denotes the end-over-end rotation). We are left with 6 lines of (yet) unknown origin, all located within 13\,\AA\ from the main laser line (see Fig.~\ref{fig:rot}). 

\section{Discussion}\label{sec:discussion}

Up until now, the majority of observations of the up-link beams from laser guide-star systems at astronomical observatories would have resulted from the un-intentional \textit{collision} between neighboring telescopes: an observational nuisance against which dedicated coordination tools were promptly developed \citep[][]{Summers2003,Summers2006,Amico2010,Amico2015}. The fact remains, however, that propagating a laser beam through the atmosphere entails a range of physical phenomena, some of which are unrelated to the creation of artificial reference wavefront sources in support of AO systems. Raman scattering, in particular, implies that initially-monochromatic up-link laser beams possess a rich spectral signature, with numerous lines located several tens (for pure rotational Raman lines) up to several hundreds (for ro-vibrational Raman lines) of \AA\ away from the main laser line. This may drive specific design choices for the next generation of optical AO instruments relying on laser guide-star systems. Under specific circumstances, for example with an on-axis laser guide-star system, pure rotational Raman lines more than one order of magnitude brighter than sky lines may potentially be found $\geq$50\,\AA\ away from the main laser line, as illustrated in Fig.~\ref{fig:rovib}. This might require the use of notch filters with a specific spectral width, should a cost-benefit analysis (balancing the scientific benefit(s) of increased wavelength coverage with the spectral contamination of rotational Raman lines) warrant it. 

The inelastic Raman scattering of laser photons provides an ideal means to characterize the new ESPRESSO spectrograph at the VLT, and validate the wavelength solution of its data product over a spectral range of $\sim$1000\,\AA. The possible presence of wind along the line-of-sight (and thus the spectral shifting of the signal over time) implies that the Raman lines are not suitable to characterize the goal \textit{precision} of 10\,cm\,s$^{-1}$ over 10 years for radial velocity measurements with ESPRESSO. We find, however, that the accuracy with which the Raman lines ($<$5\,fm for the case of $^{16}$O$_2$; see the Supplemental Material for details) and the exciting laser wavelength ($<$11.6\,fm at the 3$\sigma$ level) are known is ideally matched to the target \textit{accuracy} of ESPRESSO of 10\,m\,s$^{-1}\equiv$20\,fm at the 4LGSF laser wavelength. The 4LGSF lasers have a spectral full-width at half maximum $\lesssim$5\,MHz$\equiv$6\,fm, so that the full-width at half maximum of the observed Raman lines is actually dominated by thermal broadening (of the order of 2.3\,GHz for $^{14}$N$_2$ and 2.1\,GHz for $^{16}$O$_2$ at 273\,K). We find that measuring line-centroid values with an accuracy $<$10\,fm requires peak-to-noise values larger than $\sim$200. This was achieved, for the pure-rotational Raman lines of $^{14}$N$_2$ and $^{16}$O$_2$ in the ESPRESSO spectrum presented in Fig.~\ref{fig:rot}, up to $J\sim15$ for both molecules. 

The presence of wind shifts along the line of sight during such an observation can be easily minimized by pointing the telescope, on a quiet night, at a low airmass and at 90 deg from the dominant wind direction at high altitude. Doing so will restrict the spectral offsets caused by wind shifts to those associated with vertical motions, which can be expected to be less than 1\,m\,s$^{-1}\equiv$2\,fm \citep{Masciadri1999}. We voluntarily do not present here the formal characterization of the spectral accuracy of ESPRESSO. It will be discussed by the instrument Consortium in a separate publication, together with the required modeling of the instrument line-spread-function which falls outside of the scope of this article.

The growing number of laser guide-star systems in operation \citep[][]{dOrgeville2016} implies that the physics of Raman scattering is readily accessible to many professional astronomical telescopes world-wide: either directly, or via voluntary laser \textit{collisions} with neighboring facilities (at sites hosting multiple observatories). As the power of new generations of laser guide-star systems increases, so does the flux of the associated Raman lines, which reduces the amount of time required to acquire spectra with sufficient signal-to-noise for an accurate instrumental characterization. The molecular physics potentially accessible via Raman scattering is undeniably very rich, as evidenced by its extensive use for atmospheric studies \citep{Leonard1967,Cooney1968,Melfi1969,Melfi1972,Cooney1972,Keckhut1990,Whiteman1992,Heaps1996,Behrendt2002,Wandinger2005}. For astronomers, the $^{14}$N$_2$ and $^{16}$O$_2$ molecules are undoubtedly the best molecules to focus on. As the prime constituents of Earth's atmosphere, these homonuclear diatomic molecules will provide the strongest Raman signal, which is nowadays very well characterized. In essence, our ESPRESSO observation of one 4LGSF up-link laser beam paints the picture of a future in which laser guide-star systems at astronomical observatories will not only be thought of as mere sub-components of complex adaptive optics systems, but also as bright, flexible, and accurate calibration sources for characterizing new generations of astronomical spectrographs.

$ $\newline
\begin{acknowledgments}
\vspace{-15pt}
{\smaller \noindent We thank Ronald Holzl\"ohner for enlightening conversations, Jorge Lillo-Box and \'Alvaro Ribas for sharing their expertise in MCMC techniques with us, Susana Cerda and Rodrigo Romero for their operational support during part of the ESPRESSO commissioning run of February 2018, and the AOF Builders \citep{Arsenault2013} for the construction and installation of the AOF on UT4 at the VLT. We are very grateful to the ESO librarians for the outstanding support of our bibliographic excursions beyond the astrophysical realm. This research has made use of the following \textsc{python} packages: \textsc{matplotlib} \citep{Hunter2007}, \textsc{astropy}, a community-developed core \textsc{python} package for Astronomy \citep{AstropyCollaboration2013, AstropyCollaboration2018}, \textsc{aplpy}, an open-source plotting package for \textsc{python} \citep{Robitaille2012}, \textsc{fcmaker} \citep{Vogt2018a,Vogt2018b}, a \textsc{python} module to create ESO-compliant finding charts for OBs on \textit{p2}, \textsc{astroquery}, a package hosted at \url{https://astroquery.readthedocs.io} which provides a set of tools for querying astronomical web forms and databases \citep{Ginsburg2017}, \textsc{astroplan} \citep{Morris2018}, and \textsc{emcee} \citep{Foreman-Mackey2013}. This research has also made use of the \textsc{aladin} interactive sky atlas \citep{Bonnarel2000}, of \textsc{saoimage ds9} \citep{Joye2003} developed by Smithsonian Astrophysical Observatory, and of NASA's Astrophysics Data System. Portions of this article present research carried out at the Jet Propulsion Laboratory, California Institute of Technology, under contract with the National Aeronautics and Space Administration. Government sponsorship is acknowledged.\\ The Portuguese contribution of this work was supported by the Science and Technology Foundation FCT/MCTES through national funds and by FEDER - Fundo Europeu de Desenvolvimento Regional through COMPETE2020 - Programa Operacional Competitividade e Internacionalização through research grants: UID/FIS/04434/2019; PTDC/FIS-AST/32113/2017 \& POCI-01-0145-FEDER-032113; PTDC/FIS-AST/28953/2017 \& POCI-01-0145-FEDER-028953; PTDC/FIS-AST/29245/2017; PTDC/FIS-AST/1526/2014. Based on observations made with ESO Telescopes at the La Silla Paranal Observatory under Program ID 60.A-9128(C). All the observations described in this article are freely available online from the ESO Data Archive. }
\end{acknowledgments}

\appendix
\section{\larger\larger \sc Supplementary Material}
\section{Rotational and ro-vibrational Raman line wavelengths for non-rigid singlet diatomic rotators}\label{sec:theory_FVo}
The vibro-rotational energy of a molecule, $E_n(J)$, for given vibrational quantum number $n$ and rotational quantum number $J$, can be expressed as:
\begin{equation}
E_n(J) = G_n + F_n(J),
\end{equation}
with $G_n$ the vibrational component, and $F_n(J)$ the rotational component. Following \cite{Herzberg1950}, for diatomic molecules like $^{14}$N$_2$ with a $^{1}\Sigma$ electronic ground state or $^{16}$O$_2$ with a non-singlet electronic state but unresolved line splitting structure: 
\begin{equation}
G_n \cong \omega_e(n+\frac{1}{2}) - \omega_eX_e(n+\frac{1}{2})^2 + \omega_eY_e(n+\frac{1}{2})^3 + \omega_eZ_e(n+\frac{1}{2})^4, 
\end{equation}
and
\begin{equation}
F_n(J) \cong B_nJ(J+1) - D_nJ^2(J+1)^2,
\end{equation}
where
\begin{eqnarray}
B_n &\cong& B_e - \alpha_e(n+\frac{1}{2}) +\gamma_e(n+\frac{1}{2})^2\\
D_n &\cong& D_e + \beta_e(n+\frac{1}{2})
\end{eqnarray}
The adopted form of $F_n(J)$ assumes that molecules are non-rigid rotators, and includes the vibrational stretching of the molecular bond driven by rotation. The parameters $\omega_e$, $\omega_eX_e$, $\omega_eY_e$, $\omega_eZ_e$, $B_e$, $\alpha_e$, $\gamma_e$, $D_e$ and $\beta_e$ are molecular constants typically expressed in cm$^{-1}$ (see Table~\ref{table:params_FVo}). We adopt the values of \cite{Shimauchi1995} for $^{16}$O$_2$, \cite{Orlov1997} for $^{14}$N$_2$, \cite{Bendtsen2001} for $^{14}$N$^{15}$N, and \cite{Huber1979} for $^{14}$N$_2^+$.

\begin{table*}[htb!]
\caption{Molecular parameters adopted in our analytical model of non-rigid singlet diatomic rotators.}\label{table:params_FVo}
\begin{tabular}{c c c c c c c c c c}
\hline
\hline
&$\omega_e$ & $ \omega_eX_e$ &$\omega_eY_e$ & $\omega_eZ_e$ & $B_e$ & $\alpha_e$ & $\gamma_e$ & $D_e$ & $\beta_e$ \\
& [cm$^{-1}$] & [cm$^{-1}$] & [10$^{-2}$ cm$^{-1}$] & [10$^{-3}$ cm$^{-1}$] & [cm$^{-1}$] & [10$^{-2}$ cm$^{-1}$] &[10$^{-5}$ cm$^{-1}$] & [10$^{-6}$ cm$^{-1}$]& [10$^{-8}$ cm$^{-1}$]\\ [1ex]
\hline
$^{16}$O$_2$ &1580.161$^{a}$ & 11.95127$^{a}$ & 4.58489$^{a}$ & -1.87265$^{a}$ & 1.44562$^{a}$ &  1.59305$^{a}$ & 0 &  4.839$^{b}$ & 0 \\
$^{14}$N$_2$ &2358.54024$^{c}$   & 14.30577$^{c}$     &-0.50668$^{c}$    &  -0.1095$^{c}$      & 1.9982399$^{c}$ & 1.731281$^{c}$    &-2.8520$^{c}$ &5.7376$^{c}$    & 1.02171$^{c}$ \\
$^{14}$N$^{15}$N & - & - & - & - & 1.93184882$^{d}$ & 1.646624$^{d}$ & -2.22$^{d}$ & 5.3477$^{d}$ & 1.06$^{d}$ \\
$^{14}$N$_2$$^{+}$ & - & - & - & - & 1.93176$^{b}$ & 1.881$^{b}$ & 0 & 6.10$^{b}$ & 0 \\
\hline
\hline
\end{tabular}
\footnotetext{From \cite{Shimauchi1995}.} 
\footnotetext{From \cite{Huber1979}.}
\footnotetext{From \cite{Orlov1997}.}
\footnotetext{From \cite{Bendtsen2001}.}
\end{table*}

The frequency shift of pure rotational Raman lines (with respect to the exciting frequency) associated with the elastic scattering of photons by non-rigid singlet diatomic molecules in their ground vibrational state ($n=0$), separated in the Stokes and anti-Stokes branches, can thus be expressed as: 
\begin{eqnarray}
\Delta\nu_0|_\text{Stokes}(J) &=&E_0(J+2)-E_0(J)\\
&=&(4B_0-6D_0)(J+\frac{3}{2}) - 8D_0(J+\frac{3}{2})^3\nonumber
\end{eqnarray}
for $J=0,1,\ldots$, and:
\begin{eqnarray}
\Delta\nu_0|_\text{anti-Stokes}(J) &=&E_0(J-2)-E_0(J)\\
&=&(4B_0-6D_0)(J-\frac{1}{2}) - 8D_0(J-\frac{1}{2})^3\nonumber
\end{eqnarray}
for $J=1,2,\ldots$. The frequency shift (with respect to the exciting frequency) associated with the ro-vibrational Raman lines, separated in the $Q$, $S$ and $O$ branches, can be expressed as:
\begin{eqnarray}
\Delta\nu_{1\leftarrow0}|_\text{Q}(J) &=&E_1(J)-E_0(J)\text{ for }J=0,1,\ldots\\
\Delta\nu_{1\leftarrow0}|_\text{S}(J) &=&E_1(J+2)-E_0(J)\text{ for }J=0,1,\ldots\nonumber\\
\Delta\nu_{1\leftarrow0}|_\text{O}(J) &=&E_1(J-2)-E_0(J)\text{ for }J=2,3,\ldots\nonumber
\end{eqnarray}
For $^{16}$O$_2$, only odd values of $J$ are allowed. In our notation, the \textit{Raman shift} of a given diatomic molecule, a notion often employed in the literature, corresponds to:
\begin{eqnarray}
\Delta\nu_{1\leftarrow0}|_\text{Q}(J=0) &=& G_1 - G_0\\
& = & \omega_e - 2\omega_eX_e+\frac{13}{4}\omega_eY_e+5\omega_eZ_e, \nonumber
\end{eqnarray}
corresponding to 2329.913 cm$^{-1}$ for $^{14}$N$_2$ and 1556.398 cm$^{-1}$ for $^{16}$O$_2$, given the molecular constants of Table~\ref{table:params_FVo}.

\section{Pure rotational Raman line wavelengths of $^{16}$O$_2$}\label{sec:theory_SYu}

The spectral resolution of ESPRESSO is such that the pure rotational fine-structure lines of $^{16}$O$_2$ are resolved in the 4LGSF up-link laser beam spectrum presented in the main article. To compute the theoretical wavelength of these lines, $^{16}$O$_2$ cannot be simply treated as a non-rigid singlet diatomic rotator, given the fact that this molecule has a $^{3}\Sigma$ electronic ground state. Due to the interaction of the electronic spin \textbf{S}$=1$ and end-over-end rotational angular momentum \textbf{N}, the rotational levels of $^{16}$O$_2$ are split into three fine-structure components, corresponding to the three ways of combining \textbf{S} and \textbf{N} vectorially to form the total angular momentum \textbf{J}. Each of these fine-structure levels is labelled with quantum numbers $N$ and $J$, with $J=N-1$, $N$, and $N+1$. Once again, the nuclear spin statistics of $^{16}$O$_2$ only allows odd values of $N$. 

\cite{Yu2014} published the energies for all the fine-structure levels of $^{16}$O$_2$. To identify the associated lines in the spectrum acquired with ESPRESSO, we computed the pure rotational Raman shifts directly from these energy levels, with the following selection rules:
\begin{equation}
\Delta N=0,\pm2\qquad \Delta J=0,\pm1,\pm2
\end{equation}

The derived Raman shifts and predicted wavelengths for the pure rotational Raman lines of $^{16}$O$_2$ are presented in Table~\ref{table:params_SYu}. Each rotational transition is labelled as $^{\Delta N}\Delta J(J,N)$, with the usual conventions that $Q\equiv(\Delta=0)$, $R\equiv(\Delta=+1)$, $P\equiv(\Delta=-1)$, $S\equiv(\Delta=+2)$ and $O\equiv(\Delta=-2)$. The strongest rotational transitions occur for $\Delta N=\Delta J$, while for $\Delta N=\Delta J\pm1$ and $\Delta N=\Delta J\pm2$, the transitions scale as $N^{-1}$ and $N^{-3}$, respectively. One should note that the derived rotational Raman shifts have an accuracy $<$0.0001\,cm$^{-1}\equiv4$\,fm at the 4LGSF laser wavelength, resulting from an extensive Hamiltonian model that was used to simultaneously fit the microwave, THz, infrared, visible and ultraviolet transitions of all six oxygen isotopologues \citep{Yu2012,Drouin2012,Drouin2013}. In Table~\ref{table:params_SYu}, wavelengths are purposely rounded to two digits (equivalent to a pm level) for simplicity. Readers interested in more precise values should derive them from the associated Raman shift $\Delta\nu$ via:
\begin{eqnarray}
\lambda_\text{Stokes} &=& c\cdot\left(\frac{c}{\lambda_\text{laser}} - \Delta\nu\right)^{-1},\text{ and}\\
\lambda_\text{anti-Stokes} &=& c\cdot\left(\frac{c}{\lambda_\text{laser}} + \Delta\nu\right)^{-1},
\end{eqnarray}
with $c$ the speed of light.

In the ESPRESSO spectrum presented in the main article, the transitions in the immediate vicinity of the laser wavelength belong to the $\Delta N=0$, $\Delta J=\pm2,$ transitions, i.e. the $^{Q}$O and $^{Q}S$ branches. The line at 5892.28\,\AA\ is a blend of all $^{Q}R$ and $^{Q}P$ transitions together with the $^{Q}S(0,1)$ line. The line at 5895.86\,\AA\ is $^{S}R(1,1)$. The line at 5896.55\,\AA\ is a blend of $^{S}S(1,1)$, $^{S}S(2,1)$ and the very weak $^{S}Q(2,1)$. The line at 5897.23\,\AA\ is a blend of $^{S}S(0,1)$ and $^{S}R(2,1)$. For $N\geq3$, the line structure consists of a strong center line with a weak satellite line at each end. The strong center lines are from the blend of the three $^{S}S(J,N)$ lines and the very weak $^{S}Q(N+1,N)$ line. The weak satellites are from $^{S}R(N,N)$ and $^{S}R(N+1,N)$, respectively. Their intensities relative to the center line decrease very rapidly with $N$: in 900\,s on-source with ESPRESSO, we detected the weak satellite lines up to $N=9$. 

\LTcapwidth=\columnwidth
\begin{longtable}{c @{(} c @{,} c@{)\  } c c c}
\caption{Theoretical $^{16}$O$_2$ pure rotational Raman shifts and associated Stokes and anti-Stokes line wavelengths, computed from the energies of the molecule's fine-structure levels derived by \cite{Yu2014}, for the main exciting wavelength of the 4LGSF lasers.}\label{table:params_SYu}\\
\hline\hline
$^{\Delta N}\Delta J$ & $J$ & $N$ & $\Delta\nu$ & anti-Stokes & Stokes \\
\multicolumn{1}{c}{} & \multicolumn{1}{c}{} & \multicolumn{1}{c}{} & \multicolumn{1}{c}{[cm$^{-1}$]} & \multicolumn{1}{c}{[\AA]} & \multicolumn{1}{c}{[\AA]}  \\
\hline
\endfirsthead

\caption{continued.}\\
\hline\hline
$^{\Delta N}\Delta J$ & $J$ & $N$ & $\Delta\nu$ & anti-Stokes & Stokes \\
\multicolumn{1}{c}{} & \multicolumn{1}{c}{} & \multicolumn{1}{c}{} & \multicolumn{1}{c}{[cm$^{-1}$]} & \multicolumn{1}{c}{[\AA]} & \multicolumn{1}{c}{[\AA]}
\\ \hline
\endhead

\hline
\endfoot

$^{Q}S$ & 4 & 5 & 0.0239 & 5891.58 & 5891.60 \\ 
$^{Q}O$ & 8 & 7 & 0.0424 & 5891.58 & 5891.61 \\ 
$^{Q}O$ & 10 & 9 & 0.0943 & 5891.56 & 5891.62 \\ 
$^{Q}S$ & 2 & 3 & 0.1347 & 5891.54 & 5891.64 \\ 
$^{Q}O$ & 12 & 11 & 0.1397 & 5891.54 & 5891.64 \\ 
$^{Q}O$ & 14 & 13 & 0.1816 & 5891.53 & 5891.65 \\ 
$^{Q}O$ & 16 & 15 & 0.2213 & 5891.51 & 5891.67 \\ 
$^{Q}O$ & 18 & 17 & 0.2597 & 5891.50 & 5891.68 \\ 
$^{Q}O$ & 20 & 19 & 0.2971 & 5891.49 & 5891.69 \\ 
$^{Q}O$ & 22 & 21 & 0.3338 & 5891.48 & 5891.71 \\ 
$^{Q}O$ & 24 & 23 & 0.3701 & 5891.46 & 5891.72 \\ 
$^{Q}O$ & 26 & 25 & 0.4059 & 5891.45 & 5891.73 \\ 
$^{Q}O$ & 28 & 27 & 0.4415 & 5891.44 & 5891.74 \\ 
$^{Q}O$ & 30 & 29 & 0.4768 & 5891.43 & 5891.76 \\ 
$^{Q}O$ & 32 & 31 & 0.5120 & 5891.41 & 5891.77 \\ 
$^{Q}O$ & 34 & 33 & 0.5470 & 5891.40 & 5891.78 \\ 
$^{Q}O$ & 36 & 35 & 0.5818 & 5891.39 & 5891.79 \\ 
$^{Q}O$ & 38 & 37 & 0.6166 & 5891.38 & 5891.81 \\ 
$^{Q}O$ & 40 & 39 & 0.6513 & 5891.36 & 5891.82 \\ 
$^{Q}O$ & 42 & 41 & 0.6860 & 5891.35 & 5891.83 \\ 
$^{Q}O$ & 44 & 43 & 0.7206 & 5891.34 & 5891.84 \\ 
$^{Q}O$ & 46 & 45 & 0.7551 & 5891.33 & 5891.85 \\ 
$^{Q}O$ & 48 & 47 & 0.7896 & 5891.32 & 5891.86 \\ 
$^{Q}O$ & 50 & 49 & 0.8241 & 5891.31 & 5891.88 \\ 
$^{Q}O$ & 52 & 51 & 0.8586 & 5891.29 & 5891.89 \\ 
$^{Q}O$ & 54 & 53 & 0.8930 & 5891.28 & 5891.90 \\ 
$^{Q}O$ & 56 & 55 & 0.9274 & 5891.27 & 5891.91 \\ 
$^{Q}O$ & 58 & 57 & 0.9619 & 5891.26 & 5891.93 \\ 
$^{Q}O$ & 60 & 59 & 0.9963 & 5891.24 & 5891.94 \\ 
$^{Q}O$ & 62 & 61 & 1.0307 & 5891.23 & 5891.95 \\ 
$^{Q}O$ & 64 & 63 & 1.0652 & 5891.22 & 5891.96 \\ 
$^{Q}R$ & 64 & 65 & 1.4478 & 5891.09 & 5892.09 \\ 
$^{Q}R$ & 62 & 63 & 1.4645 & 5891.08 & 5892.10 \\ 
$^{Q}R$ & 60 & 61 & 1.4812 & 5891.08 & 5892.10 \\ 
$^{Q}R$ & 58 & 59 & 1.4980 & 5891.07 & 5892.11 \\ 
$^{Q}R$ & 56 & 57 & 1.5147 & 5891.06 & 5892.12 \\ 
$^{Q}R$ & 54 & 55 & 1.5315 & 5891.06 & 5892.12 \\ 
$^{Q}R$ & 52 & 53 & 1.5482 & 5891.05 & 5892.13 \\ 
$^{Q}R$ & 50 & 51 & 1.5651 & 5891.05 & 5892.13 \\ 
$^{Q}R$ & 48 & 49 & 1.5819 & 5891.04 & 5892.14 \\ 
$^{Q}R$ & 46 & 47 & 1.5988 & 5891.04 & 5892.15 \\ 
$^{Q}R$ & 44 & 45 & 1.6156 & 5891.03 & 5892.15 \\ 
$^{Q}R$ & 42 & 43 & 1.6326 & 5891.02 & 5892.16 \\ 
$^{Q}R$ & 40 & 41 & 1.6496 & 5891.02 & 5892.16 \\ 
$^{Q}R$ & 38 & 39 & 1.6666 & 5891.01 & 5892.17 \\ 
$^{Q}R$ & 36 & 37 & 1.6836 & 5891.01 & 5892.18 \\ 
$^{Q}R$ & 34 & 35 & 1.7008 & 5891.00 & 5892.18 \\ 
$^{Q}R$ & 32 & 33 & 1.7180 & 5890.99 & 5892.19 \\ 
$^{Q}R$ & 30 & 31 & 1.7352 & 5890.99 & 5892.19 \\ 
$^{Q}R$ & 28 & 29 & 1.7526 & 5890.98 & 5892.20 \\ 
$^{Q}R$ & 26 & 27 & 1.7701 & 5890.98 & 5892.21 \\ 
$^{Q}R$ & 24 & 25 & 1.7878 & 5890.97 & 5892.21 \\ 
$^{Q}R$ & 22 & 23 & 1.8056 & 5890.96 & 5892.22 \\ 
$^{Q}R$ & 20 & 21 & 1.8236 & 5890.96 & 5892.22 \\ 
$^{Q}R$ & 18 & 19 & 1.8420 & 5890.95 & 5892.23 \\ 
$^{Q}R$ & 16 & 17 & 1.8607 & 5890.94 & 5892.24 \\ 
$^{Q}P$ & 2 & 1 & 1.8768 & 5890.94 & 5892.24 \\ 
$^{Q}R$ & 14 & 15 & 1.8801 & 5890.94 & 5892.24 \\ 
$^{Q}R$ & 12 & 13 & 1.9003 & 5890.93 & 5892.25 \\ 
$^{Q}R$ & 10 & 11 & 1.9217 & 5890.92 & 5892.26 \\ 
$^{Q}R$ & 8 & 9 & 1.9455 & 5890.92 & 5892.27 \\ 
$^{Q}P$ & 4 & 3 & 1.9496 & 5890.91 & 5892.27 \\ 
$^{Q}R$ & 6 & 7 & 1.9735 & 5890.91 & 5892.28 \\ 
$^{Q}P$ & 6 & 5 & 1.9877 & 5890.90 & 5892.28 \\ 
$^{Q}R$ & 4 & 5 & 2.0116 & 5890.89 & 5892.29 \\ 
$^{Q}P$ & 8 & 7 & 2.0159 & 5890.89 & 5892.29 \\ 
$^{Q}P$ & 10 & 9 & 2.0398 & 5890.88 & 5892.30 \\ 
$^{Q}P$ & 12 & 11 & 2.0614 & 5890.88 & 5892.31 \\ 
$^{Q}P$ & 14 & 13 & 2.0818 & 5890.87 & 5892.31 \\ 
$^{Q}S$ & 0 & 1 & 2.0843 & 5890.87 & 5892.31 \\ 
$^{Q}R$ & 2 & 3 & 2.0843 & 5890.87 & 5892.31 \\ 
$^{Q}P$ & 16 & 15 & 2.1014 & 5890.86 & 5892.32 \\ 
$^{Q}P$ & 18 & 17 & 2.1204 & 5890.85 & 5892.33 \\ 
$^{Q}P$ & 20 & 19 & 2.1391 & 5890.85 & 5892.33 \\ 
$^{Q}P$ & 22 & 21 & 2.1575 & 5890.84 & 5892.34 \\ 
$^{Q}P$ & 24 & 23 & 2.1756 & 5890.84 & 5892.35 \\ 
$^{Q}P$ & 26 & 25 & 2.1937 & 5890.83 & 5892.35 \\ 
$^{Q}P$ & 28 & 27 & 2.2116 & 5890.82 & 5892.36 \\ 
$^{Q}P$ & 30 & 29 & 2.2294 & 5890.82 & 5892.36 \\ 
$^{Q}P$ & 32 & 31 & 2.2472 & 5890.81 & 5892.37 \\ 
$^{Q}P$ & 34 & 33 & 2.2649 & 5890.81 & 5892.38 \\ 
$^{Q}P$ & 36 & 35 & 2.2826 & 5890.80 & 5892.38 \\ 
$^{Q}P$ & 38 & 37 & 2.3003 & 5890.79 & 5892.39 \\ 
$^{Q}P$ & 40 & 39 & 2.3179 & 5890.79 & 5892.40 \\ 
$^{Q}P$ & 42 & 41 & 2.3355 & 5890.78 & 5892.40 \\ 
$^{Q}P$ & 44 & 43 & 2.3531 & 5890.77 & 5892.41 \\ 
$^{Q}P$ & 46 & 45 & 2.3708 & 5890.77 & 5892.41 \\ 
$^{Q}P$ & 48 & 47 & 2.3884 & 5890.76 & 5892.42 \\ 
$^{Q}P$ & 50 & 49 & 2.4060 & 5890.76 & 5892.43 \\ 
$^{Q}P$ & 52 & 51 & 2.4236 & 5890.75 & 5892.43 \\ 
$^{Q}P$ & 54 & 53 & 2.4413 & 5890.74 & 5892.44 \\ 
$^{Q}P$ & 56 & 55 & 2.4589 & 5890.74 & 5892.44 \\ 
$^{Q}P$ & 58 & 57 & 2.4766 & 5890.73 & 5892.45 \\ 
$^{Q}P$ & 60 & 59 & 2.4943 & 5890.73 & 5892.46 \\ 
$^{Q}P$ & 62 & 61 & 2.5120 & 5890.72 & 5892.46 \\ 
$^{Q}P$ & 64 & 63 & 2.5297 & 5890.71 & 5892.47 \\ 
$^{Q}R$ & 0 & 1 & 3.9611 & 5890.22 & 5892.97 \\ 
$^{S}R$ & 1 & 1 & 12.2918 & 5887.33 & 5895.86 \\ 
$^{S}Q$ & 2 & 1 & 14.1686 & 5886.68 & 5896.51 \\ 
$^{S}S$ & 2 & 1 & 14.3033 & 5886.63 & 5896.56 \\ 
$^{S}S$ & 1 & 1 & 14.3761 & 5886.60 & 5896.59 \\ 
$^{S}S$ & 0 & 1 & 16.2529 & 5885.95 & 5897.24 \\ 
$^{S}R$ & 2 & 1 & 16.2529 & 5885.95 & 5897.24 \\ 
$^{S}R$ & 3 & 3 & 23.8629 & 5883.32 & 5899.89 \\ 
$^{S}Q$ & 4 & 3 & 25.8125 & 5882.65 & 5900.56 \\ 
$^{S}S$ & 4 & 3 & 25.8364 & 5882.64 & 5900.57 \\ 
$^{S}S$ & 3 & 3 & 25.8745 & 5882.62 & 5900.59 \\ 
$^{S}S$ & 2 & 3 & 25.9473 & 5882.60 & 5900.61 \\ 
$^{S}R$ & 4 & 3 & 27.8241 & 5881.95 & 5901.27 \\ 
$^{S}R$ & 5 & 5 & 35.3953 & 5879.33 & 5903.90 \\ 
$^{S}S$ & 6 & 5 & 37.3406 & 5878.66 & 5904.58 \\ 
$^{S}S$ & 5 & 5 & 37.3688 & 5878.65 & 5904.59 \\ 
$^{S}Q$ & 6 & 5 & 37.3830 & 5878.64 & 5904.60 \\ 
$^{S}S$ & 4 & 5 & 37.4069 & 5878.64 & 5904.60 \\ 
$^{S}R$ & 6 & 5 & 39.3565 & 5877.96 & 5905.28 \\ 
$^{S}R$ & 7 & 7 & 46.9115 & 5875.35 & 5907.92 \\ 
$^{S}S$ & 8 & 7 & 48.8331 & 5874.69 & 5908.59 \\ 
$^{S}S$ & 7 & 7 & 48.8570 & 5874.68 & 5908.60 \\ 
$^{S}S$ & 6 & 7 & 48.8850 & 5874.67 & 5908.61 \\ 
$^{S}Q$ & 8 & 7 & 48.9274 & 5874.66 & 5908.62 \\ 
$^{S}R$ & 8 & 7 & 50.8729 & 5873.98 & 5909.30 \\ 
$^{S}R$ & 9 & 9 & 58.4156 & 5871.38 & 5911.94 \\ 
$^{S}S$ & 10 & 9 & 60.3156 & 5870.73 & 5912.60 \\ 
$^{S}S$ & 9 & 9 & 60.3373 & 5870.72 & 5912.61 \\ 
$^{S}S$ & 8 & 9 & 60.3610 & 5870.71 & 5912.62 \\ 
$^{S}Q$ & 10 & 9 & 60.4553 & 5870.68 & 5912.65 \\ 
$^{S}R$ & 10 & 9 & 62.3771 & 5870.02 & 5913.32 \\ 
$^{S}R$ & 11 & 11 & 69.9076 & 5867.43 & 5915.96 \\ 
$^{S}S$ & 12 & 11 & 71.7875 & 5866.78 & 5916.61 \\ 
$^{S}S$ & 11 & 11 & 71.8079 & 5866.77 & 5916.62 \\ 
$^{S}S$ & 10 & 11 & 71.8294 & 5866.76 & 5916.63 \\ 
$^{S}Q$ & 12 & 11 & 71.9690 & 5866.72 & 5916.68 \\ 
$^{S}R$ & 12 & 11 & 73.8693 & 5866.06 & 5917.34 \\ 
$^{S}R$ & 13 & 13 & 81.3867 & 5863.48 & 5919.98 \\ 
$^{S}S$ & 14 & 13 & 83.2472 & 5862.84 & 5920.63 \\ 
$^{S}S$ & 13 & 13 & 83.2668 & 5862.83 & 5920.64 \\ 
$^{S}S$ & 12 & 13 & 83.2870 & 5862.82 & 5920.64 \\ 
$^{S}Q$ & 14 & 13 & 83.4685 & 5862.76 & 5920.71 \\ 
$^{S}R$ & 14 & 13 & 85.3486 & 5862.11 & 5921.37 \\ 
$^{S}R$ & 15 & 15 & 92.8515 & 5859.54 & 5924.00 \\ 
$^{S}S$ & 16 & 15 & 94.6932 & 5858.90 & 5924.64 \\ 
$^{S}S$ & 15 & 15 & 94.7123 & 5858.90 & 5924.65 \\ 
$^{S}S$ & 14 & 15 & 94.7316 & 5858.89 & 5924.66 \\ 
$^{S}Q$ & 16 & 15 & 94.9529 & 5858.81 & 5924.73 \\ 
$^{S}R$ & 16 & 15 & 96.8137 & 5858.18 & 5925.39 \\ 
$^{S}R$ & 17 & 17 & 104.3004 & 5855.61 & 5928.02 \\ 
$^{S}S$ & 18 & 17 & 106.1237 & 5854.98 & 5928.66 \\ 
$^{S}S$ & 17 & 17 & 106.1424 & 5854.98 & 5928.67 \\ 
$^{S}S$ & 16 & 17 & 106.1612 & 5854.97 & 5928.67 \\ 
$^{S}Q$ & 18 & 17 & 106.4208 & 5854.88 & 5928.76 \\ 
$^{S}R$ & 18 & 17 & 108.2628 & 5854.25 & 5929.41 \\ 
$^{S}R$ & 19 & 19 & 115.7317 & 5851.69 & 5932.04 \\ 
$^{S}S$ & 20 & 19 & 117.5370 & 5851.07 & 5932.67 \\ 
$^{S}S$ & 19 & 19 & 117.5553 & 5851.07 & 5932.68 \\ 
$^{S}S$ & 18 & 19 & 117.5737 & 5851.06 & 5932.69 \\ 
$^{S}Q$ & 20 & 19 & 117.8708 & 5850.96 & 5932.79 \\ 
$^{S}R$ & 20 & 19 & 119.6944 & 5850.34 & 5933.43 \\ 
$^{S}R$ & 21 & 21 & 127.1436 & 5847.79 & 5936.06 \\ 
$^{S}S$ & 22 & 21 & 128.9310 & 5847.18 & 5936.69 \\ 
$^{S}S$ & 21 & 21 & 128.9492 & 5847.17 & 5936.69 \\ 
$^{S}S$ & 20 & 21 & 128.9673 & 5847.16 & 5936.70 \\ 
$^{S}Q$ & 22 & 21 & 129.3011 & 5847.05 & 5936.82 \\ 
$^{S}R$ & 22 & 21 & 131.1067 & 5846.43 & 5937.45 \\ 
$^{S}R$ & 23 & 23 & 138.5344 & 5843.89 & 5940.07 \\ 
$^{S}S$ & 24 & 23 & 140.3042 & 5843.29 & 5940.70 \\ 
$^{S}S$ & 23 & 23 & 140.3222 & 5843.28 & 5940.70 \\ 
$^{S}S$ & 22 & 23 & 140.3400 & 5843.28 & 5940.71 \\ 
$^{S}Q$ & 24 & 23 & 140.7101 & 5843.15 & 5940.84 \\ 
$^{S}R$ & 24 & 23 & 142.4978 & 5842.54 & 5941.47 \\ 
$^{S}R$ & 25 & 25 & 149.9023 & 5840.01 & 5944.09 \\ 
$^{S}S$ & 26 & 25 & 151.6545 & 5839.42 & 5944.71 \\ 
$^{S}S$ & 25 & 25 & 151.6724 & 5839.41 & 5944.71 \\ 
$^{S}S$ & 24 & 25 & 151.6900 & 5839.40 & 5944.72 \\ 
$^{S}Q$ & 26 & 25 & 152.0959 & 5839.27 & 5944.86 \\ 
$^{S}R$ & 26 & 25 & 153.8661 & 5838.66 & 5945.49 \\ 
$^{S}R$ & 27 & 27 & 161.2453 & 5836.15 & 5948.10 \\ 
$^{S}S$ & 28 & 27 & 162.9801 & 5835.56 & 5948.71 \\ 
$^{S}S$ & 27 & 27 & 162.9980 & 5835.55 & 5948.72 \\ 
$^{S}S$ & 26 & 27 & 163.0155 & 5835.55 & 5948.72 \\ 
$^{S}Q$ & 28 & 27 & 163.4569 & 5835.40 & 5948.88 \\ 
$^{S}R$ & 28 & 27 & 165.2096 & 5834.80 & 5949.50 \\ 
$^{S}R$ & 29 & 29 & 172.5618 & 5832.30 & 5952.10 \\ 
$^{S}S$ & 30 & 29 & 174.2793 & 5831.71 & 5952.71 \\ 
$^{S}S$ & 29 & 29 & 174.2971 & 5831.71 & 5952.72 \\ 
$^{S}S$ & 28 & 29 & 174.3144 & 5831.70 & 5952.73 \\ 
$^{S}Q$ & 30 & 29 & 174.7913 & 5831.54 & 5952.89 \\ 
$^{S}R$ & 30 & 29 & 176.5265 & 5830.95 & 5953.51 \\ 
$^{S}R$ & 31 & 31 & 183.8498 & 5828.46 & 5956.11 \\ 
$^{S}S$ & 32 & 31 & 185.5501 & 5827.88 & 5956.71 \\ 
$^{S}S$ & 31 & 31 & 185.5678 & 5827.88 & 5956.72 \\ 
$^{S}S$ & 30 & 31 & 185.5851 & 5827.87 & 5956.72 \\ 
$^{S}Q$ & 32 & 31 & 186.0971 & 5827.70 & 5956.90 \\ 
$^{S}R$ & 32 & 31 & 187.8150 & 5827.11 & 5957.51 \\ 
$^{S}R$ & 33 & 33 & 195.1076 & 5824.64 & 5960.10 \\ 
$^{S}S$ & 34 & 33 & 196.7907 & 5824.07 & 5960.70 \\ 
$^{S}S$ & 33 & 33 & 196.8084 & 5824.06 & 5960.71 \\ 
$^{S}S$ & 32 & 33 & 196.8256 & 5824.05 & 5960.71 \\ 
$^{S}Q$ & 34 & 33 & 197.3725 & 5823.87 & 5960.91 \\ 
$^{S}R$ & 34 & 33 & 199.0733 & 5823.29 & 5961.51 \\ 
$^{S}R$ & 35 & 35 & 206.3332 & 5820.83 & 5964.09 \\ 
$^{S}S$ & 36 & 35 & 207.9992 & 5820.27 & 5964.69 \\ 
$^{S}S$ & 35 & 35 & 208.0168 & 5820.26 & 5964.69 \\ 
$^{S}S$ & 34 & 35 & 208.0340 & 5820.26 & 5964.70 \\ 
$^{S}Q$ & 36 & 35 & 208.6158 & 5820.06 & 5964.90 \\ 
$^{S}R$ & 36 & 35 & 210.2995 & 5819.49 & 5965.50 \\ 
$^{S}R$ & 37 & 37 & 217.5248 & 5817.04 & 5968.08 \\ 
$^{S}S$ & 38 & 37 & 219.1738 & 5816.48 & 5968.66 \\ 
$^{S}S$ & 37 & 37 & 219.1914 & 5816.48 & 5968.67 \\ 
$^{S}S$ & 36 & 37 & 219.2085 & 5816.47 & 5968.68 \\ 
$^{S}Q$ & 38 & 37 & 219.8251 & 5816.26 & 5968.90 \\ 
$^{S}R$ & 38 & 37 & 221.4917 & 5815.70 & 5969.49 \\ 
$^{S}R$ & 39 & 39 & 228.6807 & 5813.27 & 5972.05 \\ 
$^{S}S$ & 40 & 39 & 230.3126 & 5812.72 & 5972.63 \\ 
$^{S}S$ & 39 & 39 & 230.3302 & 5812.71 & 5972.64 \\ 
$^{S}S$ & 38 & 39 & 230.3472 & 5812.71 & 5972.65 \\ 
$^{S}Q$ & 40 & 39 & 230.9986 & 5812.49 & 5972.88 \\ 
$^{S}R$ & 40 & 39 & 232.6481 & 5811.93 & 5973.47 \\ 
\hline
\end{longtable}

\bibliographystyle{apsrev4-1_lim10}
\nocite{apsrev41Control}
\bibliography{revtex4-1_prx-hack,bibliography_fixed}

\end{document}